# Utilization of Skin Color Change for Image-based Tactile Sensing

Seitaro Kaneko[1], Hiroki Ishizuka[2], Hidenori Yoshimura[3], and Hiroyuki Kajimoto[1]

[1]Department of Informatics, The University of Electro-Communication, 1-5-1 Chofugaoka, Chofu, Tokyo 182-8585, Japan

[2]Graduate School of Engineering Science, Osaka University, 1-3 Machikaneyama, Toyonaka, Osaka 560-8531, Japan

[3]Department of Engineering and Design, Kagawa University, 2217-20 Hayashi-cho, Takamatsu, Kagawa 761-0396, Japan

**Abstract—** Measurement of pressure distribution applied to a fingertip is crucial for the teleoperation of robots and human computer interface. Previous studies have acquired pressure distribution by affixing a sensor array to the fingertip or by optically recording the deformation of an object. However, these existing methods inhibit the fingertip from directly contacting the texture, and the pressure applied to the fingertip is measured indirectly. In this study, we propose a method to measure pressure distribution by directly touching a transparent object, focusing on the change in skin color induced by the applied pressure, caused by blood flow. We evaluated the relationship between pressure and skin color change when local pressure is applied, and found a correlation between the pressure and the color change. However, the contact area and the color change area did not align perfectly. We further explored the factor causing the spatial non-uniformity of the color change, by accounting for the stress distribution using finite element analysis. These results suggest that the proposed measurement method can be utilized to measure the internal stress distribution, and it is anticipated to serve as a simple sensor in the field of human computer interface.

## 1. Introduction

Humans perceive the shape of grasping objects, such as bricks, pencils, and scissors, using the pressure distribution on the skin [1]–[3]. Research is underway to develop and use sensors to record the pressure distribution information applied to the human skin. This technology is indispensable for recording, reproducing, and sharing tactile information in the field of tele-robotics and human computer interaction. In these studies, arrays of strain gauges, conductive rubber, and capacitance sensors are typically employed to measure the pressure distribution when the fingertip contacts the sensors. Furthermore, high spatial resolution can be achieved by applying technologies such as micro electro mechanical systems.

However, existing sensors must be placed between the contact object and the fingertip because direct pressure needs to be applied to the sensing elements. The configuration of the sensors inhibits a user from directly touching the object. To cope with this issue, methods have been proposed that abandon measuring the spatial distribution of force and instead focus on measuring only the magnitude of the force. For example, a method was proposed to estimate the contact force using skin deformation on the side of the finger [4]. Tanaka et al. used vibration propagation to calculate the contact force [5]. Moreover, Mascaro et al. [6]–[8] proposed a method to estimate the fingertip force using the fingertip nail's color change caused by changes in the blood flow. Their method can effectively assess the magnitude and direction of vertical and shear forces applied to the finger. Although the above methods can estimate the contact force applied to the finger without covering the objects or finger skin, they cannot measure the pressure distribution on the skin's surface.

We propose a method to directly measure the change in skin color (blanching) on the contact surface by using a transparent object and a camera underneath it to estimate the pressure and the corresponding distribution (Figure 1). The change in skin color is caused by the change in blood flow due to the external force and captured using a camera. The proposed method eliminates a sensor between the skin and the object. The main limitation of this method is that the contact object must be transparent and require a built-in camera. However, it has a potential application in the situation where the contacting object can be prepared beforehand, such as human computer interface. Furthermore, camera-based high-resolution measurement of pressure distribution may contribute to the understanding of skin contact mechanism.

To realize the proposed method, it is necessary to confirm that the pressure distribution corresponds to the color change distribution on the skin. Previous studies have reported the correspondence between the amount of color change and the input pressure at a specific point [9] [10]. This study evaluates the relationship between pressure distribution and skin color change distribution when a bump shape is pressed against the skin. We first investigated whether the area of contact coincides with the area of color change, and whether there is a correlation between the pressure value presented to the finger and the skin color change. Subsequently, we investigated whether the stress distribution inside the skin could explain the skin color change, using finite element method (FEM) analysis, to explore the factor causing the spatial discrepancy between the contact shape and the color change.

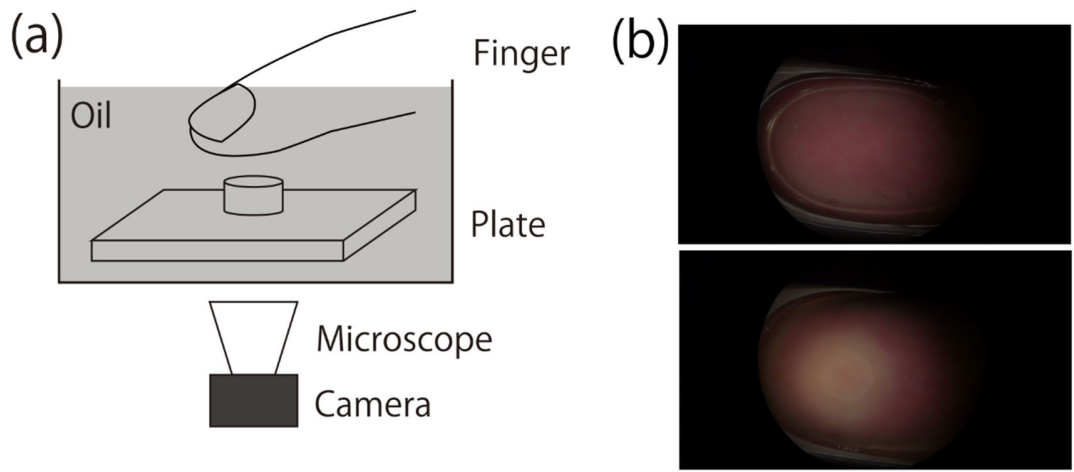

Figure 1 (a) Schematic diagram of the measurement system. A transparent object is submerged in oil to suppress optical refraction, and the image of the fingertip is captured by using a microscope. (b) Examples of actual measurement. Images show before (top) and after (bottom) pressing a finger against the convexity. Skin color change is observed when a finger is pressed down on the convex (in this case, a cylinder).

## 2. Related work
### 2.1 Basic skin structure

The skin structure is described in a three-layer model of the epidermis, dermis, and subcutaneous tissue. Each layer contains tactile receptors, which are sensory nerves that are receptive to touch. It has been suggested that these layers have a complex structure, which is considered to increase the sensitivity of the tactile receptors. For example, a computer simulation showed that strain energy is concentrated near the dermal papilla, where Merkel cells exist between the epidermis and the dermis by shear and compression [11].

In the skin, a group of fine blood vessels called capillaries exists, of which there are three main areas: In the papillary dermis between the epidermis and the dermis, in the upper dermis, and at the boundary between the subcutaneous tissue and the dermis [12]. These capillaries are 0.07 mm, 0.3 mm, and 1 mm deep from the skin surface [13]. Capillaries near the dermal papilla run perpendicular to the skin surface, while the shallow and deep dermis capillary network runs tangential to the skin surface. Moreover, blood does not flow in the epidermis, the skin's surface. A skin model for simulating the multilayered structure of the skin, including the arrangement of such capillaries, has been proposed, primarily for measuring the light propagation characteristics inside the skin [13], [14].

### 2.2   Skin color change

The skin changes its surface color under external pressure. The color change phenomenon is called the blanching. It has been demonstrated that the blanching is caused by pressure-induced obstruction of blood flow in the capillaries in the skin area, which inhibits the absorption of green light by hemoglobin. This color change occurs when a pressure of 5.5 kPa (42 mmHg) or greater is exerted on the skin [15]. The color changes as the exerted pressure increases [16], [17]. It has been reported that such color change is caused by vertical pressure and skin shear [18]. The skin color change is typically observed at joints owing to finger bending and stretching [9].

Measurement techniques using such color changes are primarily used in medical applications. For example, such techniques are used to determine bedsores and to measure the pulse rate. When bedsores occur, hemoglobin leaks out of the capillaries, so the color does not change even when an external pressure is applied. This feature is used as a medical technique to visually determine bedsores by applying pressure to the target area via a transparent plate [19], [20]. Moreover, measuring heart rate and blood pressure can be performed non-invasively by irradiating intense light from the outside and measuring weak color changes [10], [21]. These technologies have already been implemented in smartphones and smartwatches, and their use continues to expand.

The measurement technique that uses color change is also used for tactile sensing. A typical example is a method proposed by Mascaro et al. [6] for sensing the direction of force applied to the fingertip by measuring the color change on the fingernail. Their method enables wearable force measurement without placing a sensor between the contact object and the fingertip's surface. Such research has also been actively conducted in recent years, including the measurement method of photographing the color deformation of fingernails using a camera [22], [23] and the measurement of color deformation on fingernails on a viscous object [24]. However, methods that directly exploit the color change occurring on the skin have yet to be investigated. In particular, the spatial distribution characteristics of color change occurring on the skin have yet to be investigated.

## 2.3 Skin simulation

Owing to the mechanically complex structure of the fingertip, the stress distribution affecting the contact conditions on the skin surface and the blood flow occurring in the finger may be different. However, measuring the stress distribution inside the finger is difficult and needs simulation. Simulation models, primarily represented by the finite element method analysis, have been used to determine the stress distribution occurring inside the skin from the contact conditions on the skin surface.

In the early stages of research, various methods have been proposed [25], [26]. Serina et al. [25] modeled the fingertip as an elliptical membrane, and Srinivasan et al. [26] modeled the skin tissue as a two-dimensional (2D) infinite surface. Recent developments in finite element method analysis that better describe the nonlinear properties of the fingertip have been used. For example, Maeno et al. [11] measured the mechanical properties of the components of the finger, such as the epidermis and dermis. They developed a model on a computer based on these results. Dandekar et al. [27] established a 2D finite element method analysis model simulating the skin structure. They showed a correlation between the strain energy applied to mechanoreceptors in the skin and the frequency of nerve activity. When a rectangular convexity was pressed against the fingertip, the strain energy was observed to be concentrated near the convex edge.

## 3. Experimental method

In the experiments, we investigate whether the contact area coincides with the area of color change and if there's a correlation between applied pressure and color change. Additionally, we used finite element methodanalysis to examine if the stress distribution inside the skin could explain the spatial roll of the color change.

## 3.1    Stimulus

Cylindrical bump textures were used for both actual measurement and computational simulation. Eight conditions were prepared, with cylinder diameters of 2, 3, 4, and 5 mm and convex heights of 1 and 2 mm. The stimuli for actual measurement were fabricated by affixing the acrylic convex part to an acrylic plate using adhesive.

## 3.2 Apparatus

The skin deformation measurement system developed by Tanaka et al. [28] was utilized, with the number of cameras reduced from two to one. The fingertips were magnified by a stereo microscope installed in the lower part of the oil tank, and the images were recorded using a camera (RX0, SONY) (Figure 2 (a)). Measurements were performed using only one side of the binocular microscope, and the images were captured at 60 frames per second. Participants were asked to align their fingers with a guide and press down. Measurements were performed while the participants touched a texture submerged in silicone oil (KF-56A, Shin-Etsu Chemical Co., Ltd.) stored in an oil tank. The oil has a refractive index of 1.498, equivalent to the refractive index of 1.490 of the acrylics prepared as the sample, thus eliminating any hindrance to observation due to optical refraction and scattering. A force plate (T F-2020, Tech Gihan) was used to measure the pressing force (Figure 2 (b)). A laboratory jack was used to position the participant's arm at the top of the device.

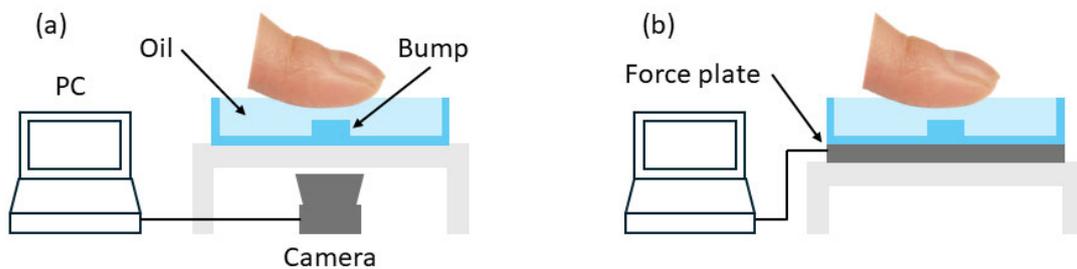

Figure 2 Schematic illustration of experimental setups. (a) Skin color measurement. (b) Force measurement.

### 3.3 Participants

Eight participants (22-26 years old, all male, one left-handed and seven right-handed) used their dominant hand for the experiment. This study was conducted with the approval of the Ethics Committee of the University of Electro-Communications (Approval No. 20010).

### 3.4 Procedure

First, the participant's hands were washed and inspected to ensure there is no scratch on the fingertips as preparation for the experiment. Second, the participants practiced the pressing procedure to guarantee the amount of pressing down would be equivalent to the convex thickness of the texture. Specifically, the participants were instructed to press down their fingers so that their fingertips touch the convexity of the texture and the base of the texture just barely not contact the skin, thus to ensure that the height of the convexity becomes equal to the dent of the skin. Before starting the measurement of each texture, the position of the base was adjusted so that the center of the convexity of the texture remained at a fixed position when the image was captured. This was done to extract the averages for each participant for later analysis. At the beginning of each trial, the participants held the finger surface floating in oil without touching the convex surface to record the initial state before the color change occurred. A beeping sound was emitted 4 seconds from the start of measurement, and the participant pressed his fingertip against the convexity. The duration of pressing was 3 seconds. During the trial, the elapsed time from the start of the measurement and the image being captured by the camera could be viewed on the monitor.

These procedures were performed for each texture five times for the skin color change measurement using the camera and three times for the pressure measurement using the force plate. Thus, the total numbers of trials per participant was 40 (= 4 (width) x 2 (height) x 5 (repetitions)) for the skin color change measurement and 24 (= 4 (width) x 2 (height) x 3 (repetitions)) for the pressure measurement. The order of the trials for the conditions was in the descending order of the diameter of the 2 mm thick texture and descending order of the diameter of the 1 mm thick texture.

### 3.5 Analysis

We conducted finite FEM analysis to analyze the experimental results. The proposed method is based on the compression of blood vessels caused by internal pressure within the skin. We discuss this point by comparing the actual measurement with the FEM analysis.

The following procedure was employed to extract the amount and area of color change on the skin (Figure 3). First, the green channel of the RGB video was extracted. This is because hemoglobin, which is responsible for the color change, absorbs green component well [9], [16], [29]. Subsequently, 100 frames were extracted from the noncontact and contact images, and the average of each image was calculated. The difference between the average image during the contact and the noncontact image was then calculated, and this was used as the observed amount of color change. One-dimensional (1D) color change was plotted every 20 degrees and averaged to estimate the relationship between the distance and color change. The above process was performed for each condition, and the average value for a participant was used as the result.

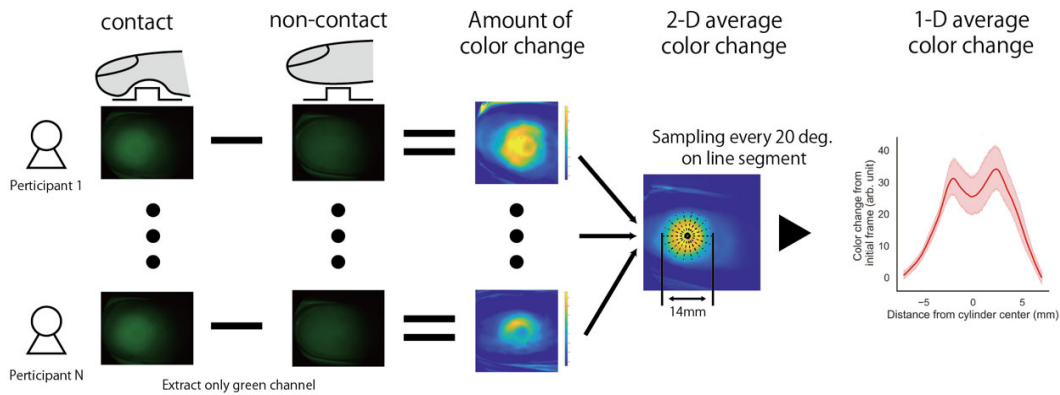

Figure 3 Analysis procedure for measured skin color change. The green channel is extracted from the video, and the 100-frame average of each pressing and non-pressing phase is extracted. The average amount of color change computed for each participant is then extracted. It is defined as the amount of 2D color change. Subsequently, resampling is performed on a 14 mm line segment passing through the convex center, and the average of the extracted results is plotted every 20 degrees. This is defined as the amount of one-dimensional (1D) color change. This is extracted for later use in finite element method analysis results.

### 3.6 Simulation

A finite element method analysis was performed under the same conditions as the skin color measurement to discuss the experimental result—the fingertip model in the finite element method analysis is illustrated in Figure 4 (a). The model reproduced five types of fingertip tissue elements: epidermis, dermis, subcutaneous tissue, bone, and nail. A skin displacement of 2 mm is assumed to be a linear elastic body [11]. In this study, following previous research, we also assumed the finger to be a linear elastic body. The parameters were determined based on related studies [30], [31] (Table 1). Fingerprint tissue was formed on the contact area. The distance between fingerprints was 0.35 mm, and the convexity was 0.1 mm. In the simulation, a rigid model that emulated an acrylic cylinder model was placed on the fingertip model, as depicted in Figure 4 (b). The model did not emulate the flat plate portion because only the convex portion was touched in the measurement experiment. Since we considered the contact between skin and acrylic, the friction coefficient between the fingertip and the rigid model was 0.4 [32]. The rigid body reached the desired displacement in 1 s and then held it for 3 s. The time increment for finite element method analysis was set to 1 ms. Von Mises stress at the final calculation step was sampled. The width of the rigid model was 2, 3, 4, and 5 mm, and the model's displacement was 1 and 2 mm. In total, eight calculations were conducted. In the subsequent analysis, the sampled von Mises stresses were replotted on a 2D grid from the mesh points to regress the internal stress distribution. The extracted region was 7 mm wide from the center of the convexity to be pressed and 2 mm deep from the skin surface. Commercially available software for finite element method analysis (Marc 2021, MSC Software) was used to perform the simulation.

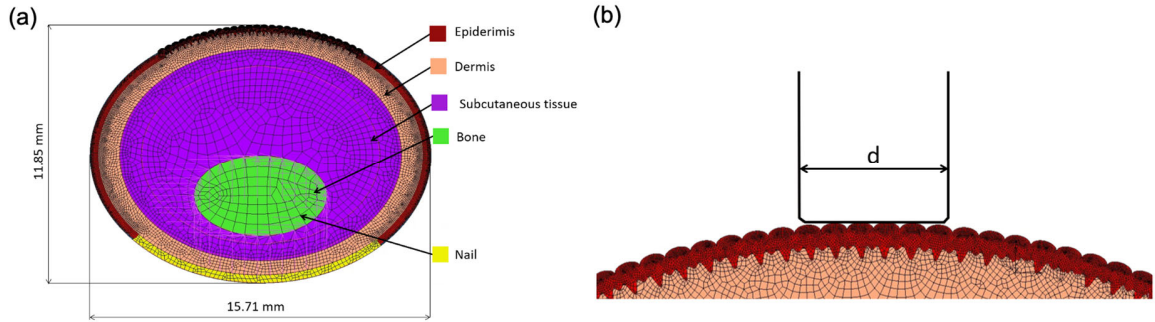

Figure 4 finite element method analysis model of a human fingertip. (a) The cross-sectional model that consists of the epidermis, the dermis, subcutaneous tissue, bones, and nails. The parameters for each layer were determined from the related studies. [30], [31] (b) The skin was indented using a rigid model with a width of d (2, 3, 4, 5) mm that emulated the acrylic cylinder used in the experiment.

Table 1 Skin parameters used in the finite element method analysis.

| Layer | Elastic modulus (Mpa) | Poison ratio |
| --- | --- | --- |
| Epidermis | 0.136 | 0.48 |
| Dermis | 0.080 | 0.48 |
| Subcutaneous tissue | 0.034 | 0.48 |
| Bone | 17000 | 0.30 |
| Nail | 170.000 | 0.30 |

## 3.7 Regression between experiment and simulation

Our hypothesis is that the color changes observed at the skin surface can be explained by the weighted addition of strains at different skin depths. Based on this idea, weight parameters for skin depth were estimated by linear regression to explain the correspondence between the calculated internal skin stress distribution and the actual measured skin color change. This enables us to determine 1) whether the internal stress distribution can explain the distribution of color change and 2) which skin depth region is the main contributor to skin color change.

Partial Least Squares regression was employed. This was because the strain values for each depth extracted by finite element method were so similar (correlation was so high) that it was impossible to perform a normal multiple regression. In the regression, the 1D data of color change sampled from the skin color change measurements were used as the objective variable, and the depth of the stress distribution by finite element method analysis was used as the explanatory variable. In this study, all measured data were used to ensure that the correspondence was established. Accordingly, regressions were performed for each condition. These regressions used the mean of the color change result of each participant.

## 4. Experimental result
### 4.1 Skin color change

Skin color change was depicted when a convex surface was pressed (Figure 5 (a)). The yellow area in the figure indicates that the difference between the green channel component has increased between before and after pressing. The convex edge existed in the area indicated by the black dotted circle in the resulting image. 1D skin color change was extracted from the 2D image and plotted as a graph to observe the color change in the convexity. The area indicated in gray in the graph is the same as the width of the convexity (Figure 5 (b)). The result showed that skin color change occurred in the area where the fingertip was in contact with the convexity. It is evident that as the bump width increased, the area of apparent color change shifted from the entire bump to the convex edge. In particular, at 4 and 5 mm bumps, the color change was more evident at the edges than at the center of the bump. This indicates that the contact area and the color change distribution do not coincide perfectly.

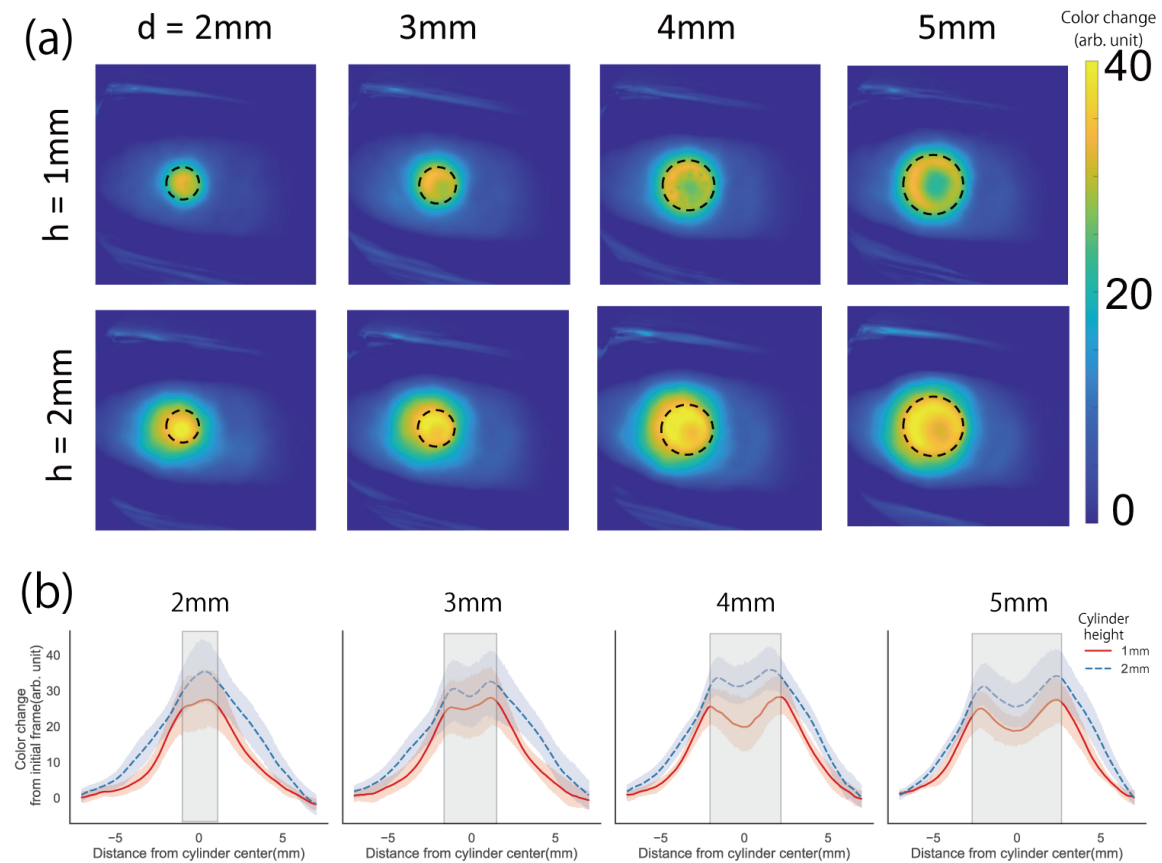

Figure 5 Average skin color change volume for all participants. (a) Inter-participant average of the amount of skin color change considered. Brighter color indicates more significant change. The black dotted lines indicate the convex edge of the sample. (b) 1D graph of color change. The area in gray represents the contact area.

Furthermore, we analyzed whether the pressure on the contact surface corresponded to the amount of skin color change. The averaged pressure is presented in Figure 6. The averaged pressure values were largely different between cases with a convexity width of 2 mm and other cases, but the changes in pressure were relatively small in the other cases. This result may be because the pressing force increased with the increase of the convexity area. The relationship between the pressure and the amount of skin color change for each participant is depicted in Figure 7. The horizontal axis represents the averaged pressure, and the vertical axis represents the average color change within the contact area. Consequently, although large individual differences were observed in the pressure and the color change, the amount of color change increased as the pressure increased. One-dimensional linear regression was performed for each participant's results, and the mean coefficient of determination was 0.439. The analysis of variance showed that a significant difference was observed in 3 of the 8 participants at the 5% level, and an important trend at the 10% level was observed in 3 of the 8 participants. The result show that the amount of color change generated by each participant is different.

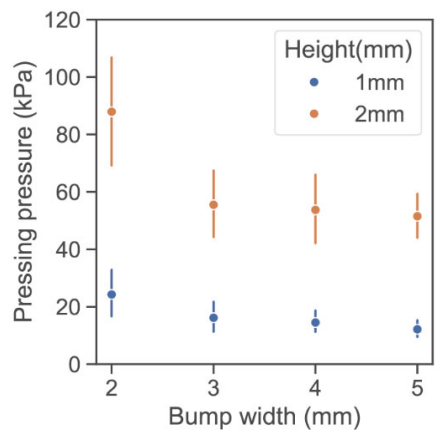

Figure 6 Pressure applied to the finger under each convexity condition. The horizontal axis shows the convexity width and the vertical axis shows the applied pressure. Each convexity height is color-coded. Error bars indicate standard error. This figure shows that higher pressure is applied to the fingertip when the convexity is high or the width of the convexity is small.

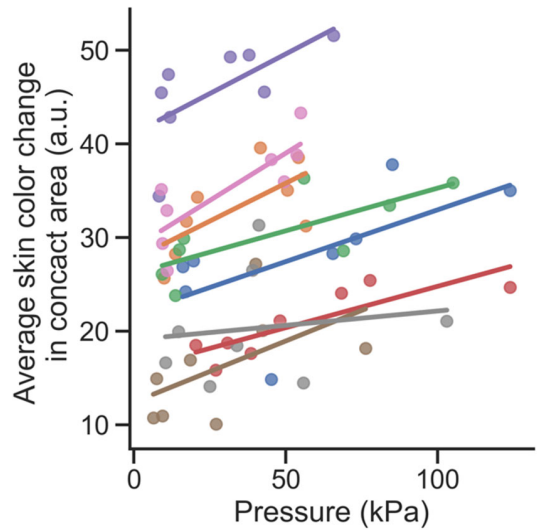

Figure 7 Relationship between the surface average pressure value per participant and the average amount of color change. The results for each participant are color-coded. Lines in the graph indicate the result of regression from the data using the least squares method. This figure shows that there were large individual differences in pressure and color change, but the amount of color change increased as the pressure increased.

### 4.2 Simulated pressure distribution inside skin

From the previous result, it was observed that the area of skin color change was not identical to the contact area but concentrated near the convex edge. It can be attributed to decreased blood flow due to stresses applied to the inside of the skin [16]. We examine whether von Mises stress, a scalar quantity that is often applied to evaluate skin deformation, can explain the amount of color change occurring on the skin surface. In this section, the result of the internal stress distribution by finite element method simulation is presented.

The result of the internal skin stress distribution extracted by finite element method was re-sampled at 1 mm intervals in both depth and width from the mesh points. The result plotted on a 2D grid are presented in Figure 8. The upper part of each figure represents the skin surface. The result indicates that a large amount of stress is observed on the skin surface in the area corresponding to the bump width. Additionally, high stress with the same width of the bump occurred at a depth deeper than 1.5 mm. These characteristics did not change even if the amount of indentation was varied. Meanwhile, the magnitude of the stress generated was apparent at the skin surface and in areas deeper than 1.5 mm.

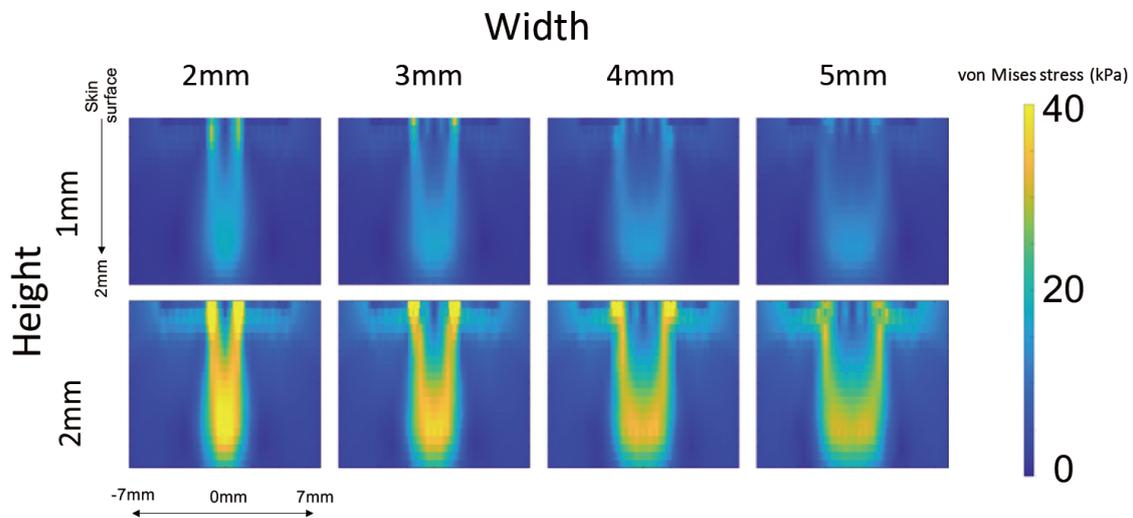

Figure 8 Stress distribution inside the skin calculated by finite element method. Yellow indicates a larger strain occurred.

## 4.3 Correspondence between skin color and stress

Based on the results, we observe the correspondence between the skin color change and the distribution of internal skin stresses. First, partial least squares regression was performed using 1D data sampled, where the actual skin color change measurements serve as the objective variable and the depth of the stress distribution computed by finite element method serves as the explanatory variable.

The coefficients of determination $R^2$ in each condition are listed in Table 2. The result revealed that the regression coefficients for all conditions exceeded 0.85, which indicates that the regression for each condition was performed with high accuracy. The relationships between position and skin color change for the measurement and the estimated result using the model are illustrated in Figure 9. The result suggests that the stress distribution inside the skin can explain the color change near the convex edge.

Subsequently, we computed the variable importance for prediction score calculated from the regression model to determine which depths of stress strongly contributed to skin color change. Depth region with the variable importance for prediction score greater than one can be considered crucial in the regression model [33] (Figure 10). The result reveals that variable importance for prediction scores exceeded one in regions less than 0.3 mm in all conditions.

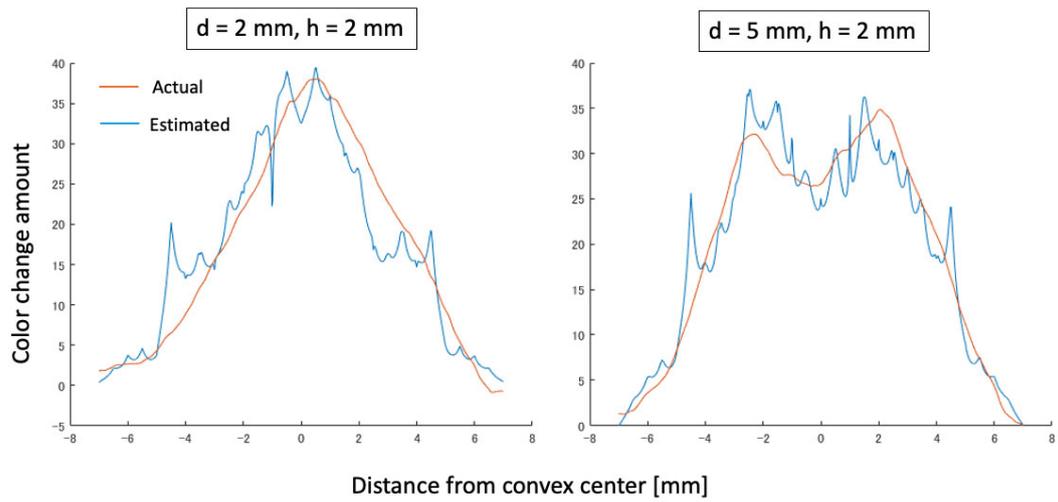
Figure 9 Relationships between distance from the convex center and the amount of color change.

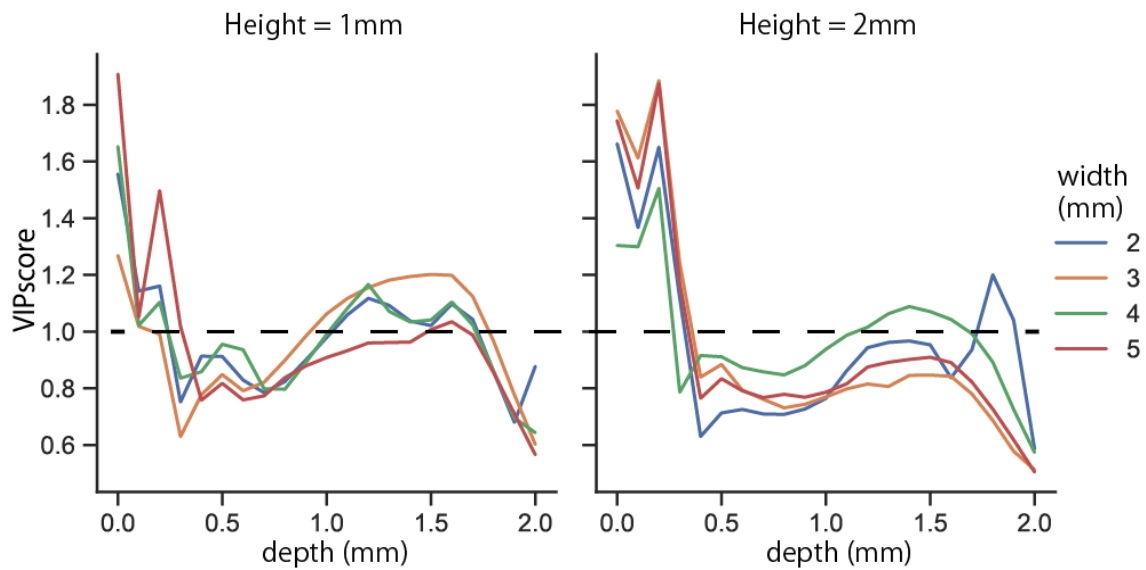

Figure 10 variable importance for prediction scores at each skin depth calculated from the partial least squares regression model. The horizontal axis represents skin depth, and the vertical axis represents the variable importance for prediction score; the score greater than one can be considered crucial for the construction of skin color changes occurring at the surface.

Table 2 Coefficients of determination from partial least squares regression for each condition. The coefficients exceeded 0.85 in all conditions, suggesting that the model was developed with high accuracy.

| Height (mm) \ width (mm) | 2 | 3 | 4 | 5 |
|---|---|---|---|---|
| 1 | 0.94 | 0.89 | 0.94 | 0.94 |
| 2 | 0.93 | 0.87 | 0.92 | 0.93 |

## 5  Discussion
### 5.1 Correspondence between skin color and stress

In this study, we proposed a method for measuring pressure distribution on fingertip by observing the color change of the skin surface when it contacts an object. To verify the proposed method, we measured the color change and pressure when a bump of varying thicknesses and diameters was pressed against a finger. The result revealed a significant color change to applied pressure distribution. The relationship between pressure and skin color change was clarified when local pressure was applied to the skin. A correlation between pressure and color change and its area was confirmed. By utilizing this correlation, it should be possible to estimate the local pressure exerted on the fingertip from the image.

Skin color changes did not occur uniformly in the contact area; instead, they occured near the bump edge. This can be attributed to the distributional blockage of blood flow within the skin. As demonstrated in previous case studies [11], [34], the stress exerted on the finger tends to be concentrated near the bump edges. The maximum value of von Mises stress in the finite element method analysis was approximately 40 kPa, which well exceeded a pressure of 4.2 kPa to occlude the capillaries [15]. Prior research also showed that the reduction in blood flow is more proportional to the pressure applied to the finger's surface [17]. This suggests that a decrease in blood flow occurred in the area where more pressure was applied, leading to the distribution of color change on the skin observed in this experiment.

We then compared the observed color change and the finite element method analysis, to see which depth of the skin contributed the color change. The result indicates that the coefficients of determination of the regression results of the color change using the stress distribution computed by our finite element method analysis are all above 0.85. It suggests that skin color change can be explained using internal skin stresses.

The result of the regression analysis also revealed that the depth regions that strongly contribute to the color change are less than 0.3 mm from the skin surface. The depth corresponds to the dermal papillae of the epidermis, where the capillary network inside the skin is present. It suggests that skin color change is caused by inhibiting blood flow to shallow area.

Interestingly, Merkel cells, which perceive fingertip pressure, reside near the dermal papilla of the epidermis. Previous studies have reported that the strain energy density amplitude at the location of Merkel cells correlates with the neural firing rate of the SAI innervating Merkel cells [11]. The result suggests that the proposed method may allow us to observe areas where Merkel cells are active.

The limitation of the model developed in this study is that it does not include factors related to light absorption and scattering due to skin properties. For instance, the amount of light penetrating the skin's interior, and its reflectance highly depends on wavelength and depth [35]. This measurement targets only green channel light for analysis. As a result, it could reduce the accuracy of the regression result because the model is constructed using only the light that appears on the skin's surface layer. To solve this problem, a simulation model of light propagation that includes blood flow inside the skin may be used. A Monte Carlo simulation model that provides for the structure of capillaries has already been developed, and by improving this model, it should be able to construct a more accurate stress color deformation model.

## 5.2 Application and limitation

The measurement of pressure distribution on the finger surface can be performed noninvasively using the method proposed in this study. Additionally, the measurement of color change can serve as an indicator to estimate the pressure applied inside the skin, as there is a correlation between the area where skin color change occurs, and the stress applied near the receptor. However, given the significant variation in the amount of color change among participants, this method is unsuitable for more precise measurements.

The proposed method has the potential to be applied as a simple and effective sensor in the field of human-computer interaction. For instance, when developing robotic sensors designed for human interaction, the contact area can be made transparent, allowing the applied load to be measured solely through captured images. A similar approach, such as GelSight [36], measures light displacement within a gel medium. In contrast, the proposed method requires only a camera, eliminating the need for additional equipment. Moreover, this technique can contribute to the design and manufacturing of wearable devices by enabling a straightforward measurement of the load exerted by a device on different parts of the human body. Specifically, a wearable device could be fabricated using transparent materials via a 3D printer, and the color variation upon wearing could be recorded using a camera. Traditional methods for assessing such loads rely on pre-simulations or high-density wearable sensors to measure forces applied to the skin. However, simulations struggle to accommodate variations in individual arm size and shape, while sensor-based methods require placing a sensor between the device and the skin, altering the actual sensation of wearing the device. In contrast, transparent contact surfaces can now be easily produced using 3D printing technology, making this method highly applicable in fabrication research and related fields.

Despite its advantages, there are several limitations when using the system as a sensor. First, it is necessary to observe the skin directly, meaning the object to be contacted must be transparent and have a built-in camera. Second, the amount of skin color change and the pressure applied to the finger vary greatly among participants. This is primarily because each participant has different skin hardness and color before pressing. Additionally, since skin color variation is heavily influenced by changes in blood flow, external factors such as temperature and emotional state may also affect measurements. For this reason, it is necessary to calibrate the correlation between the amount of skin color change and the amount of pressure in advance when building the system. Therefore, it is expected that the system can be used efficiently by combining it with a force sensor that measures the amount of pressure applied at the time of contact.

To further improve measurement accuracy, it is necessary to develop a model that estimates blood flow changes before and after compression based on skin color information. Specifically, by analyzing pre-contact images to estimate the existing blood flow and calculating the amount of blocked blood flow from the color change due to compression, the applied pressure on the fingertip can be more precisely determined. A similar model has been used to estimate pressure on the finger pad based on nail color changes [10]. By adapting this model for the fingertip, measurements can be performed independently of individual differences and environmental factors. The development of this model remains a challenge for future research.

Finally, if the surface of the contacting object has a complicated shape, it may be difficult to observe the region where internal stress is applied due to the refraction of light on the surface of the shape. To address this, potential solutions include having a transformation map that takes refraction into account when the object shape is known in advance or submerging the surface in a liquid with a refractive index like that of the material. These solutions can enhance the applicability of the method in real-world scenarios.

## 6 Conclusion

We propose a method to observe the skin's color distribution using a transparent object and a camera to measure pressure on uneven surfaces. To verify the proposed method, the correspondence between the skin color change that occurs when a finger was pressed against multiple uneven surfaces and the change in the amount of pressure applied to the fingertip was verified. As a result, the amount of skin color change increased as the pressing pressure increased. However, since the amount of color change generated varied from participant to participant, prior calibration may be necessary when the sensor is applied. In addition, it was found that the color change area was concentrated near the convex edges. To clarify the cause, we analyzed the stress distribution inside the skin during pressure presentation by simulation and verified the correspondence with the color change. The result showed that the stress distribution inside the finger can explain the color change on the finger surface with a high coefficient of determination of 0.85. Our research will serve as a base for the future development of a pressure measurement system.

## Acknowledgements

This work was supported in part by JSPS KAKENHI Grant Number JP20J23128, 20H02121 and 22H01447, Japan.

## Declaration of generative AI and AI-assisted technologies in the writing process

During the preparation of this work the authors used ChatGPT in order to translate. After using this tool, the authors reviewed and edited the content as needed and took full responsibility for the content of the publication.